\documentstyle[epsf]{aipproc}
\begin{document}
\title{Toward gravitational wave detection}
\author{L.S.~Finn$^{(1,2,3)}$, G.~Gonzalez$^{(1,2)}$, J.~Hough$^{(4)}$,
  M.F.~Huq$^{(3),(1)}$, S.~Mohanty$^{(1,2)}$, J.~Romano$^{(5)}$,
  S.~Rowan$^{(6)}$, P.R.~Saulson$^{(7)}$ and K.A.~Strain$^{(4)}$}
\address{
$^{(1)}$Center for Gravitational Physics \& Geometry, Penn State
University; 
$^{(2)}$Department of Physics, Penn State University;
$^{(3)}$Department of Astronomy and Astrophysics, Penn State University;
$^{(4)}$Department of Physics and Astronomy, The University of Glasgow;
$^{(5)}$Department of Physical Sciences, The University of Texas,
Brownsvillle;
$^{(6)}$Department of Applied Physics, Stanford University;
$^{(7)}$Department of Physics and Astronomy, Syracuse University
}
% Covers four abstracts (69, 79, 78, 90) for a total of eight pages. 
\maketitle

\begin{abstract}
  An overview of some tools and techniques being developed for data
  conditioning (regression of instrumental and environmental artifacts
  from the data channel), detector design evaluation (modeling the
  science ``reach'' of alternative detector designs and
  configurations), noise simulations for mock data challenges and
  analysis system validation, and analyses for the detection of
  gravitational radiation from gamma-ray burst sources. 
\end{abstract}

\section{Detector Characterization}\label{sec:regression}

Quality data analysis requires quality data. Part of the process of
producing quality data is identifying and, as far as possible,
removing instrumental and environmental artifacts. Here we illustrate,
using data taken during November 1994 at the LIGO~40~M prototype, the
identification and removal, through linear regression, of artifacts due
to harmonics of the 60Hz power mains.

A power spectrum (psd) of the LIGO~40~M IFO\_DMRO (interferometer
differential-mode read-out; hereafter, ``gravity-wave'') channel shows
a series of narrow spectral features at 60~Hz and its
harmonics. Similar narrow
spectral features are evident in the magnetometer channel, IFO\_MAGX,
which was recorded simultaneously with the gravity wave channel.

Focus attention, in both the magnetometer and gravity-wave channel, on
a narrow band about one of the harmonics. We suppose that, in this
narrow band, the gravity wave channel $h$ is related to the
magnetometer channel $M$ through an expression of the form
\begin{equation}
h[k] = {B(q)\over A(q)} M[k-n]  + {C(q)\over D(q)}e[k],
\end{equation}
where the index $k$ indicates the sample number, the residual $e[k]$
is white and $A(q), B(q), C(q)$ and $D(q)$ are polynomials in the lag
operator $q^{-1}$,
\begin{equation}
q^{-1}M[k] := M[k-1].
\end{equation}
The ratio $B(q)/A(q)$ is a linear filter that can be thought of as the
transfer function between the magnetometer and the gravity
wave channel; similarly, the ratio $D(q)/C(q)$ can be thought as a
filter that whitens that part of the gravity wave channel not
explained by the magnetometer channel.

Using a small sample of data we find the ``best'' filters $B(q)/A(q)$
and $C(q)/D(q)$, where better choices yield smaller residuals and have
fewer poles and zeros. (Fewer poles and zeros are desired because we
don't want to over fit the data; smaller residuals are desired because
we want to identify everything in $h$ that can be explained by $M$.)

To illustrate, we focus on the 540~Hz harmonic in an approximately
2666 second continuous stretch of LIGO 40~M data taken on 19 November
1994. We mix this harmonic down to zero frequency and down-sample the
data to a 4~Hz bandwidth. Using a $100$ second segment of data from
both the magnetometer and gravity wave channels, we find the filters
$B(q)/A(q)$ and $C(q)/D(q)$ and the lag $n$. In this case the best
filters have six zeros and one pole each. The quantity
\begin{equation}
h[k] - {B(q)\over A(q)}M[k-n]\label{eq:res}
\end{equation}
is then as free from the effects of the 540~Hz harmonic as we can make
it, under the hypotheses of this model. 

\begin{figure}
\epsfysize=0.5\textwidth
\begin{center}
\leavevmode
\epsffile{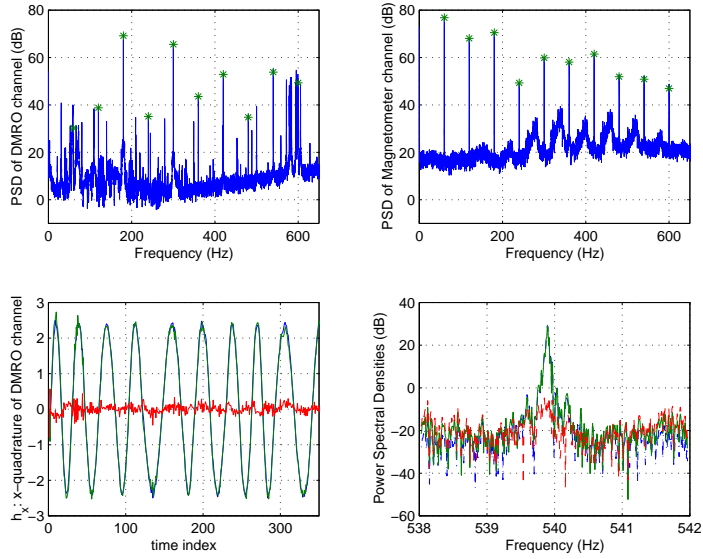}
\end{center}
\caption{60~Hz harmonics and their regression from the interferometer
  data stream. See section \ref{sec:regression} for
  details.}\label{fig:regression} 
\end{figure}

Figure \ref{fig:regression} shows the effectiveness of this analysis.
The top two panels show the gravity wave and magnetometer channel psd,
with the 60~Hz harmonic features marked with asterisks. In the left
bottom panel one curve shows one quadrature of the mixed and decimated
gravity wave channel, a second shows the prediction that comes from
applying the filter $B(q)/A(q)$ to the lagged magnetometer channel,
and the third is their difference (cf.~eq.~\ref{eq:res}). The final
panel shows the psd of this difference, superposed with the psd of the
original $h[k]$ and the magnetometer prediction. The magnetometer
channel explains 40~dB of the contamination of the gravity wave
channel by the 540~Hz harmonic.

\section{Benchmarks for detector design}\label{sec:bench}

Gravitational wave detectors are built to detect gravitational waves.
Better detectors do a better job of detecting gravitational waves.
But, what are the relative advantages of, {\em e.g.,} better
sensitivity in a narrow band as opposed to somewhat worse sensitivity,
but over a broader band? How do we quantify better?  To aid in
answering this question we have developed a Matlab model, {\tt
  bench\/}, that calculates different figures of merit, based on
source science, for use in detector design and configuration trade
studies.

An interferometer is described to {\tt bench\/} in terms its laser,
optical surface, suspension and substrate properties, since it is
these that determine the dominant contributions to the detectors noise
performance.\footnote{Gross parameters, such as arm length, may also
  be varied.}  From this characterization {\tt bench\/} determines the
detectors expected thermal noise in the mirror suspensions and
substrates, radiation pressure and laser shot noise.  Using this
idealized noise model {\tt bench\/} calculates two different figures
of merit: the first, an effective distance to which inspiraling binary
neutron stars can be observed above a fixed threshold signal-to-noise,
and the second, a measure of the upper-bound that can be placed on the
intensity of a stochastic gravitational-wave background in the LIGO
detector system, assuming identical interferometers installed at each
observatory.

The {\tt bench\/} model for the principal interferometer noise sources
has the following features:
\begin{itemize}
\item Radiation pressure and laser shot noise expressions support
  interferometer configurations including power recycling and resonant
  sideband extraction through the specification of three mirror
  transmittances and associated losses in the optical system. Thermal
  lensing effects are estimated and a warning issued if the laser
  power on the beam splitter exceeds the bounds permitted by the
  losses in the optical system.
\item The suspension thermal noise model includes thermoelastic and
  structural damping for ribbon and cylindrical suspensions composed
  of different materials (and, for ribbon suspensions, different
  aspect ratios) \cite{rowan97a,rowan97b};
\item Thermal noise in the (cylindrical) mirror substrates depends on
  substrate dimensions, material properties (Young modulus, Poisson
  ratio, loss angle), and the incident (laser) beam radius
  \cite{bondu98a}. 
\end{itemize}
  
The binary inspiral ``effective distance'' figure of merit is a
distance $r_0$ such that the observed rate of inspiraling binary
neutron star systems with S/N greater than 8 is equal to $4\pi
r_0^3\dot{n}/3$, where $\dot{n}$ is the rate per unit volume of
inspiraling binary systems \cite{finn96a}.  The stochastic signal
sensitivity benchmark determines a threshold on the cross-correlation
between two identical detectors (located and oriented like the LIGO
detectors) such that, in the absence of a stochastic signal (or any
other cross-correlated noise), the cross-correlation estimated using
1/3~yr data would exceed this threshold in only one of every one
hundred trials. Other benchmarks are planned. 

\begin{figure}
\epsfysize=0.5\textwidth
\begin{center}
\leavevmode
\epsffile{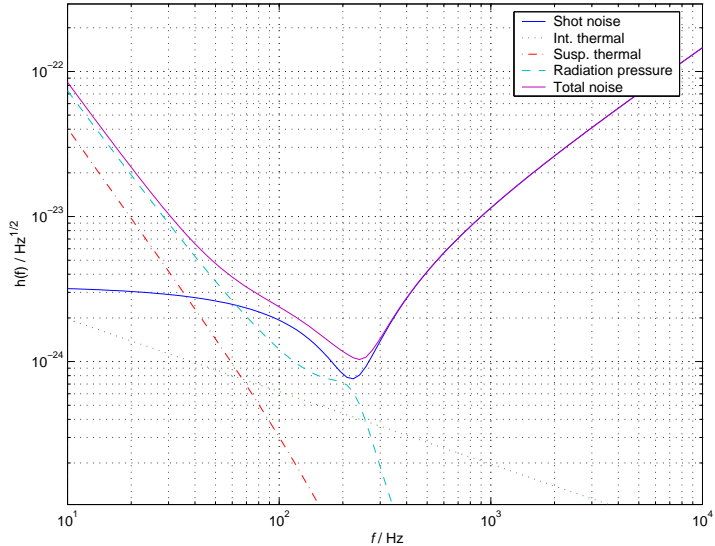}
\end{center}
\caption{A noise model produced by {\tt bench,} a tool for use in
  evaluating the science reach of different interferometric detector
  configurations or designs. An interferometer with this configuration
  can observe binary inspiral in an effective volume of radius
  288~Mpc. See section \ref{sec:bench} for details.}\label{fig:bench}
\end{figure}

Figure \ref{fig:bench} shows a sample noise model produced by bench
for an interferometer whose parameters are those described in
\cite{gustafson99a} as a possible LIGO~II goal, but whose mirror
reflectivities are optimized to maximize the distance to which neutron
star binary inspirals could be observed. In this configuration, a
single interferometer could survey an effective volume of radius
290~Mpc for neutron star binary inspirals: a volume large enough to
expect an event rate of just less than one per month. The two LIGO
interferometers operating in this configuration would observe an
effective volume $2^{1/2}$ times large, with an expected event rate of
one just less than once every two weeks.

\section{Mock data for mock data challenges}\label{sec:simdata}

Reliable analysis software is a prerequisite for reliable data
analysis. Validating the performance of analysis system software will
involve ``Mock-Data Challenges'' (MDCs). In a MDC, ``mock data'' ---
artificially generated time-series whose statistical character and
signal content is known exactly --- is passed through an analysis
pipeline. 

MDCs take two forms. In the first, idealizations of the detector
noise, for which the pipeline response can be anticipated, are
constructed and passed through the analysis pipeline. Agreement
between the anticipated and actual system response validates the
analysis system implementation. In the second form, more faithful
simulations of detector noise are used to {\em calibrate\/} the
analysis system: {\em i.e.,} determine, in a realistic but controlled
environment, the detection efficiency and false alarm frequency as a
function of the pipeline thresholds associated with the selections and
data cuts.

In either form, mock data always includes the fundamental noise
sources that contribute the greatest part of the detector noise power.
In existing and planned interferometric detectors these fundamental
contributions arise from radiation pressure noise, laser shot noise,
suspension and substrate thermal noise. The thermal noise
contributions have the character of structurally damped harmonic
oscillators with small loss angles. The significant contribution from
the substrate thermal noise arises from the low-frequency tail of the
noise distribution, whose power spectral density (psd) is proportion
to $f^{-1}$. The significant contribution from the suspension thermal
noise arises from the high-frequency tail of the suspension pendulum
mode, where the psd is proportional to $f^{-5}$. Additionally, the
resonant peaks associated with the weakly damped suspension violin
modes contribute important instrumental artifacts that must be part of
a realistic noise simulation.

The general plan of our noise simulator is to find a combination of
linear filters, acting in parallel on independent white noise
sequences, whose sum gives rise to a sequence whose power spectral
density (psd) has the desired form. The design of short, effective
linear filters that capture either the odd-power dependence on $f$
characteristic of the thermal noise tail of structurally damped
oscillators, or the strong resonant peaks of the weakly damped
systems, has been a stumbling block in this program. We have overcome
those difficulties by developing a physical model of a structurally
damped system whose noise psd has the desired in-band character.
Arising from a physical model, the psd can be factored into a real,
linear, zero-pole-gain filter that is stable and invertible ({\em
  i.e.,} has all of its poles in the left half-plane and zeros in the
right half-plane), and with the required magnitude response. The
filter's zeros, poles and gain are determined uniquely and directly by
location and quality factor of the resonance, and the desired
simulation bandwidth.

\begin{figure}
\epsfysize=0.5\textwidth
\begin{center}
\leavevmode
\epsffile{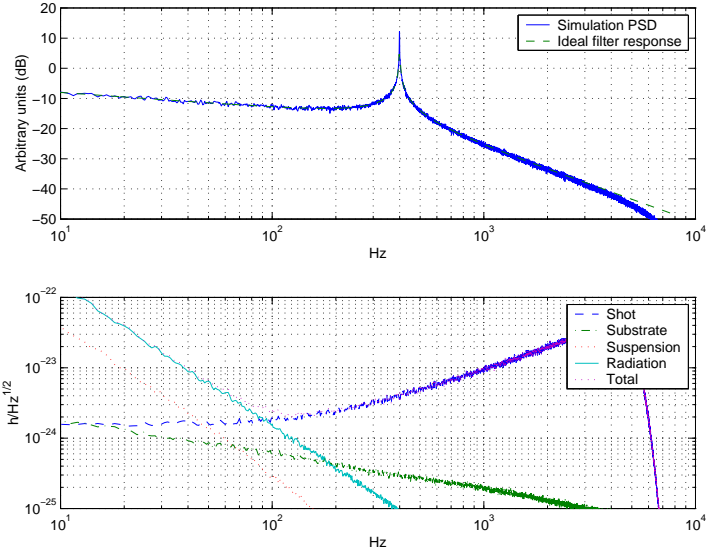}
\end{center}
\caption{Simulations of the principle contributions to the overall
  noise of an interferometric gravitational wave detector. For more
  details see section \ref{sec:simdata}.}\label{fig:simdata} 
\end{figure}

The first panel of figure \ref{fig:simdata} shows the psd of noise
simulated to have the character of a structurally damped harmonic
oscillator over three decades in frequency above and below the
resonance. The simulation psd is overlaid with the spectrum of an
idealized structurally damped harmonic oscillator with the same
loss-angle and resonance frequency. This model involved thirteen poles
and an equal number of zeros. They agree to better than 1\% over the
detector bandwidth. The second panel shows the noise psd of the other
components of the simulation (radiation pressure, shot, internal
thermal and pendulum mode suspension thermal noise) for a LIGO~II like
interferometer without resonant sideband extraction.\footnote{When
  simulating noise at the LIGO~I sample rate of 16.384~KHz the
  simulator, currently implemented as an interpreted Matlab program,
  produces mock data at the rate of 81,920 samples per second, or
  5$\times$ the real-time detectors sample rate, on a Sun Ultra-30
  workstation. The inverse is the true, not amortized, cost per
  simulated sample and holds for any number of samples.}

\section{Gravitational waves and $\gamma$-ray bursts}

Gamma-ray bursts (GRBs) are likely triggered by the violent formation
of a solar mass black hole, surrounded by a debris torus, at
cosmological distances. Given the distance, the violence of the
formation event, and the range of possible progenitors, waveforms from
events like these cannot be predicted {\em a priori,} nor the
gravitational radiation associated with an individual burst detected
directly.

Nevertheless, if GRBs are accompanied by gravitational wave bursts
(GWBs) the correlated output of two gravitational wave detectors
evaluated in the moments just prior to a GRB will differ from that
evaluated at other times. This {\em difference\/} can be detected,
with increasing sensitivity as the number of detector observations
coincident with GRBs increases. Observations at the two LIGO
observatories, operating at the anticipated LIGO~I sensitivity and
coincident with 1000 GRBs, can be used to set a 95\% confidence upper
limit of ${h}_{\text{RMS}}\sim 1.7\times10^{-22}$ on the gravitational
waves associated with GRBs. (See \cite{finn99f} for more details.)

Consider the correlation $X$ between the output $h_1$ and $h_2$ of two
LIGO gravitational wave detectors:
\begin{eqnarray}
  X&:=&\langle x_1,x_2\rangle = 
  \int\int_0^T dt\, dt^\prime\, x_1(t)
  Q(|t-t^\prime|) x_2(t^\prime), 
  \label{eq:innpro}
\end{eqnarray}
where we have adjusted the origin of time in each detector so that
plane gravitational waves from a direction $\vec{n}$ arrive
``simultaneously'' in the two detectors.  Assuming that GWB signals
from GRBs are broadband bursts, take the Fourier transform of $Q$ to
be $\widetilde{Q}(f) = \left( S_1(|f|)S_2(|f|)\right)^{-1}$, where
$S_i(f)$ is the power spectral density (psd) of detector $i$, for $f$
in the detector band, and $0$ otherwise.

Every time a GRB occurs (say, at time $t_0$) adjust the origin of time
so that $\vec{n}$ points towards the GRB and form $X$ with the
interval $(t_0-T,t_0)$ of data from the two detectors. The duration
$T$ of this interval we choose large enough so that we are likely to
have included in the interval any associated GWB. For
current models of gamma-ray bursts this is no longer than several
hundred seconds (where we have accounted for the cosmological redshift
of these distant sources). 

For each observed GRB we thus have an $X$. Collect these $X$ into the
{\em on-source\/} observation set ${\cal X}_{\rm {on}}$.  Similarly,
we build an {\em off-source\/} observation set ${\cal X}_{\rm {off}}$
following the same procedure but choosing random times $t_0$, not
associated with any GRBs, and random directions $\vec{n}$ in the sky.

Assuming that GRB signals are weak compared to the detector noise, the
sample sets ${\cal X}_{\rm off}$ and ${\cal X}_{\rm on}$ differ only
in their means.  This difference, $\overline{s}$ is just the average
over the source population of $\langle h_1, h_2\rangle$, where $h_k$
is the GWB signal in detector $k$. For the two LIGO detectors $h_1$
and $h_2$ are, to a good approximation, identical and $\overline{s}$
is proportional to the mean-square amplitude of the wave of $h$ over
the source population.

We can test for the difference in the means of the two distributions
${\cal X}_{\rm {on}}$ and ${\cal X}_{\rm {off}}$ using Student's
$t$-test \cite{snedecor67a}, a standard test for difference in means.
This provides a simple, yes/no answer to the question of whether GWBs
are associated with GRBs.

Alternatively, we can use the value of the $t$-statistic to set an
upper bound on $\overline{s}$. To assess the
strength of the upper bound, assume that there is no gravitational
radiation associated with GRBs. In this case the ensemble mean, median
and mode of the $t$ statistic is zero. Assuming that we actually
observed $t$ equal to zero we would obtain the 95\% upper bound
\begin{eqnarray}
  {h}_{\text{RMS},95\%}^2 &\leq& 
  \left[9.4\times10^{-22}\right]^2
  \left(
    {T\over500\,\text{s}}{1000\over N_{\text{on}}}
  \right)^{1/2}
  {S_0\over\left( 3\times10^{-23}\,\text{Hz}^{-1/2}\right)^2}
  \left({\Delta f\over 100\,\text{Hz}}\right)^{3/2}.
  \label{hrms1}
\end{eqnarray}
where, for convenience, we have modeled the LIGO~I detector noise as
approximately constant with power spectral density $S_0$ over the
bandwidth $\Delta f$, and much higher elsewhere. The value of $T$
adopted here is consistent with external shock models of GRBs; if, on
the other hand, it becomes clear that internal shock models are more
appropriate (as is becoming more likely), then $T$ will be reduced by
a factor of 1000 and the limit will improve by a factor of nearly 
six. 

This upper limit is remarkably strong, especially because it arises
without assuming any model for the GWB source or waveform, or the
detector noise.\footnote{The $t$ statistic is a robust one;
  correspondingly, the binary test (did we/didn't we detect) is
  insensitive to the actual distribution of the $X$ in the sets ${\cal
    X}_{\rm {on}}$ and ${\cal X}_{\rm {off}}$. Additionally, since
  each $X$ is a sum over many statistically independent random
  variables the noise contribution to each $X$ is also, by The Central
  Limits Theorem, normal.} Focusing on the difference in the
population means has the important consequence that noise correlated
between the detectors, but not associated with gravitational waves
from GRBs, does not affect the difference in the means.
Correspondingly, statistical tests built around the difference in the
means are insensitive to noise correlated between the two
gravitational wave detectors.  Observations with this sensitivity will
have important astrophysical consequences, either confirming or
constraining the black hole model for GRBs, neither or which can be
done with strictly electromagnetic observations.

\section{Acknowledgments}
We are grateful to   the LIGO Laboratory for permitting the use of
LIGO~40M prototype data in this work. 
The research described here is funded by United States National Science
Foundation awards PHY 98-00111, 99-6213, 98-00970, 96-02157, 96-30172; the
University of Glasgow and the United
Kingdom funding agency PPARC. 
%\bibliographystyle{prsty}
%\bibliography{phyjabb,references,keys}

\end{document}